\documentclass[a4paper,english]{jpconf}

\usepackage{times}
\usepackage{graphicx}
\usepackage{amssymb}

\usepackage{fancybox}
\usepackage[latin1]{inputenc}
\usepackage{babel}
\usepackage{pst-all}
\usepackage{epsfig}
\usepackage{rotating}

\newcommand{\be}{\begin{equation}}
\newcommand{\ee}{\end{equation}}
\newcommand{\bea}{\begin{eqnarray}}
\newcommand{\eea}{\end{eqnarray}}
\newcommand{\bdm}{\begin{displaymath}}
\newcommand{\edm}{\end{displaymath}}

\begin{document}

\title{What can quantum cosmology say about \\
  the inflationary universe?}

\author{Gianluca Calcagni}

\address{Instituto de Estructura de la Materia, CSIC,
calle Serrano 121, 28006 Madrid, Spain}

\author{Claus Kiefer}

\address{
Institut f\"ur Theoretische Physik, Universit\"at zu K\"oln,
 Z\"ulpicher Strasse~77, 50937 K\"oln, Germany}

\author{Christian F. Steinwachs}
\address{Physikalisches Institut, Albert-Ludwigs-Universit\"at Freiburg,
Hermann-Herder-Strasse~3, 79104 Freiburg, Germany}

\begin{abstract}
We propose a method to extract predictions from quantum cosmology for
inflation that can be 
confronted with observations. Employing the tunneling boundary
condition in quantum geometrodynamics, we derive a probability distribution for
the inflaton field.  A sharp peak in this distribution can
be interpreted as setting the 
initial conditions for the subsequent phase of inflation.
In this way, the peak sets the energy scale at which the inflationary phase has
started. This energy scale must be consistent with the energy scale
found from the inflationary potential and with the scale found from a
potential observation of primordial gravitational waves. 
Demanding a consistent history of the universe from its quantum origin to its
present state, which includes decoherence, we derive a condition that
allows one to constrain the 
parameter space of the underlying model of inflation. 
We demonstrate our method by applying it to two models: Higgs
inflation and natural inflation.
\end{abstract}


\section{Introduction}\label{Intro}

It is generally assumed that the Universe underwent a period of
quasi-exponential expansion very early in its evolution. This phase is
called {\em inflation} and has the advantage of giving a
causal explanation for structure formation (see
e.g. \cite{Mukhanov:book} for a textbook introduction). 
But while the kinematic features of inflation are well understood, its
precise dynamical origin remains unclear. There exist plenty of
different models, mostly phenomenologically devised,
and one must at present resort to observations in order to
constrain the class of allowed models \cite{Martin15}. 

How can one select inflationary models from a theoretical point of
view? The ideal situation would be to have an established fundamental
theory at one's disposal from which one can derive the dynamics of
inflation, for example in the form of an inflaton field $\varphi$ and
its potential $V(\varphi)$. Unfortunately, we do not have a theory of
this kind. This is partly related to the open question of constructing
a consistent and empirically correct quantum theory of gravity
\cite{OUP}. While the energy scale of inflation is most likely
separated from the Planck scale by some orders of magnitude, its
origin can probably not be entirely understood without reference to
quantum gravity. 

A conservative approach to quantum gravity, which should give reliable
results at least somewhat away from the Planck scale, is the direct
quantization of general relativity in standard metric variables
\cite{OUP}. This approach is called quantum geometrodynamics and has
the important feature that it leads back to general relativity in the
semiclassical limit. 
One might then use this framework to derive the desired
constraints on inflationary models. But how is this possible? 

If applied to the cosmic regime, quantum geometrodynamics leads to the
framework of quantum cosmology, with the wave function of the universe
as its central concept. This wave function obeys the Wheeler--DeWitt
equation \cite{OUP}. One could thus attempt to derive constraints on
inflation from this equation. But this is not easy, because the
interpretation of the wave function in quantum cosmology is far from
being straightforward. In our
contribution to these Proceedings, we shall review one popular method
to obtaining predictions in quantum cosmology and apply it to two
models of inflation.

In the following Sec.~\ref{PredInQC}, we shall briefly discuss how one
can envisage predictions in quantum cosmology and how one can apply
them to inflation. In Sec.~\ref{EScales}, we present the energy scales
relevant for inflation. In slow-roll models, the relevant energy scale is
derived from the corresponding potential. 
If primordial gravitational waves can be observed,
one can directly extract the inflationary energy scale, which must
then be
consistent with the scale derived from the potential. A third scale is
the one obtained from quantum cosmology, which must be consistent with
the two other scales if we adopt the criterion of selection presented
in Sec.~\ref{PredInQC}. In Sec.~\ref{SpecialModels}, we apply these
considerations to the models of non-minimal Higgs inflation and natural inflation. In this, we follow closely our earlier papers
\cite{Barvinsky:2009jd} and \cite{CKS14}, where more details and
references can be found. We end with a brief conclusion and outlook.

\section{Predictions in quantum cosmology}\label{PredInQC}

At the most fundamental level, quantum gravity --- and therefore also
quantum cosmology --- is timeless \cite{OUP,Kiefer:2013jqa}. This is a
direct consequence of the fact that general relativity does not
contain any absolute notion of time; after quantization, the dynamical
spacetime vanishes in the same way as the classical trajectories
do in quantum mechanics. In the Wheeler--DeWitt equation, there
is thus no $t$, and the notions of probability and probability
conservation (unitarity) seem to lose their usual meaning.

How, then, can one extract predictions from quantum cosmology? So far,
only heuristic ideas are available. It has been suggested that a
minimal scheme is to look for peaks in the wave
function and to interpret them as predictions, see, for example,
\cite{Hartle87,Halliwell91}. If the wave function vanishes in a
certain region of 
configuration space, this means that the corresponding values will
never occur; this property is important for the discussion of
singularity avoidance. 

Inflation is a (semi)classical concept, so a prerequisite for obtaining
inflation from quantum cosmology is an efficient quantum-to-classical
transition. This is achieved by decoherence, a process that is well
understood and experimentally explored in quantum mechanics
\cite{deco,Schlosshauer}. It has been shown that decoherence is
efficient at the ``onset of inflation'', which thus justifies the use of
robust semiclassical components of the universal wave function and the
neglect of interference terms, see, for example, \cite{OUP,Zeh07} and
the references therein. It has been suggested that one should
interpret the wave function only in the semiclassical limit
\cite{Vilenkin89}, but we leave this as an open question. For the
purpose of this contribution, we adopt the heuristic proposal that a
strong peak in the wave function is interpreted as a prediction, while we will not attempt to infer anything from it in the general case.
 In the semiclassical limit, from the wave function after
decoherence one can get also a restriction on the allowed classical trajectories, that is, one
obtains a selection criterion for trajectories. 

In the general case, one could envisage a derivation of probabilities
from quantum entanglement in the manner attempted in \cite{Zurek05}, see
also \cite{Schlosshauer}, but this is still an open issue. 

The idea to get a probability for inflation from the wave function in
this way was
entertained already in \cite{HaPa86}. A more precise formulation was
obtained in \cite{BK90} and later papers (see the references in
\cite{Barvinsky:2009jd}) by emphasizing in particular the need 
to go to the one-loop level in quantum field theory in order to obtain
a normalizable wave function. At the tree level, the slow-roll
approximation does, in general, not lead to a peak because of the
small field derivatives.     
 
Even in the semiclassical limit, the form of the wave function will
depend on the employed boundary condition. The two most popular
boundary conditions are the `no-boundary' and the `tunneling'
conditions; see, for example, \cite{OUP,Halliwell91} for detailed
reviews. In general, the no-boundary condition will not predict the
occurrence of inflation. This is why attention is concentrated on the
tunneling proposal. It should be emphasized that
`tunneling' is meant here only as a metaphor because tunneling has no
meaning in a timeless context, except in a limited sense in the
semiclassical approximation; see \cite{Zeh07,Conradi,Vilenkin15} for a
more detailed discussion. We shall also use the tunneling boundary
condition in our contribution and shall see in which sense one can get
a prediction for inflation from it. Instead of using the Wheeler--DeWitt
equation, we shall employ the equivalent path-integral formulation and
its semiclassical limit.  

The issues of probability and probability measure are even more subtle
and contrived in the recently discussed `multiverse' context (consult,
for example, 
\cite{Page} and the references therein), but this will not be addressed here.

\section{Energy scales of inflation}\label{EScales}

In this section, we will summarize how to extract in detail the energy
scale of inflation from inflationary models themselves, from
observation, and from quantum cosmological considerations.

\subsection{Slow-roll predictions for the energy scale of inflation}\label{ESlowRoll} 

For all inflationary models, the main observables are the power
spectra of primordial scalar and tensor perturbations which are
generated during inflation on top of an already existing homogeneous
and isotropic Friedmann-Robertson-Walker background space-time, 
\bea
 {\cal P}_{\rm{t}}:=A_{\rm{t}}\,\left(\frac{k}{k_{*}}\right)^{n_{\rm{t}}},\qquad {\cal P}_{\rm{s}}:=A_{\rm{s}}\,\left(\frac{k}{k_{*}}\right)^{n_{\rm{s}}-1}.\label{PowerSpectra}
\eea
The mode $k_{*}$ corresponds to a pivot scale $k_{*}^{-1}$ (to be chosen according to the observational window of the experiment) when the mode $k_{*}$ first crosses 
the Hubble scale, $k_{*} = a_{*}\, H_{*}$. Here, $H(t)=\dot{a}(t)/a(t)$ denotes the Hubble parameter.
Within the slow-roll approximation, deviations from a perfect de~Sitter stage can be parametrized to first order in terms of the two slow-roll parameters 
\bea
 \epsilon_{\rm{v}}:=\frac{M_{\rm{P}}^2}{2}\,\left(\frac{V'}{V}\right)^2,\qquad\eta_{\rm{v}}:=M_{\rm{P}}^2\,\frac{V''}{V}\,,\label{SlowRollParameters}
\eea
where $M_{\rm{P}}$ denotes the (reduced) Planck mass. 
The amplitudes of the power spectra are given by $A_{\rm{t}}$ and
$A_{\rm{s}}$. The tensor and scalar spectral indices $n_{\rm{t}}$ and
$n_{\rm{s}}$ encode the scale dependence of the power spectra (its
slope). These parameters can be entirely expressed in terms of $V$,
$\epsilon_{\rm{v}}$, and $\eta_{\rm{v}}$  
\bea
 A_{\rm{t}}=\frac{2\,V}{3\,\pi^2\,M_{\rm{P}}^4},\qquad A_{\rm{s}}=\frac{V}{24\,\pi^2\,M_{\rm{P}}^4\,\epsilon_{\rm{v}}},\qquad\label{CMBAmpl}
 n_{t}=-2\,\epsilon_{\rm{v}},\qquad n_{\rm{s}}=1+2\,\eta_{\rm{v}}-6\,\epsilon_{\rm{v}}\,.\label{Tilt}
\eea
All quantities in (\ref{PowerSpectra})--(\ref{SlowRollParameters}) must
be evaluated for the value of the inflaton field at Hubble-scale exit,
$\varphi_{*}$,
which, in turn, can be expressed in terms of the number of e-folds
$N_{*}$ by integrating and inverting the relation  
\bea
 N_{*}=\int_{t_{*}}^{t_{\rm{end}}}\,{\rm d}t\,H\simeq\frac{1}{M_{P}^2}\,\int_{\varphi_{*}}^{\varphi_{end}}\,{\rm d}\varphi\,\frac{V}{V'}\,.\label{NumberEfolds}
\eea
The value of $\varphi_{\rm{end}}$, where inflation ends, is defined by the breakdown of the slow-roll approximation, when $\epsilon_{\rm{v}}\simeq{\cal O}(1)$, which leads to the convention
\bea
\epsilon_{\rm v}(\varphi_{\rm{end}}):=1\,.\label{EndOfInfl}
\eea
The energy scale predicted by inflationary slow-roll models is then given by
\bea
E^{\rm{model}}_{\rm{infl}}:=V_{*}^{1/4}:=[V(\varphi_{*})]^{1/4}\,.
\eea

\subsection{Observational constraints for the energy scale of inflation}\label{Obs} 

The observational energy scale of inflation $E_{\rm{inf}}^{\rm{obs}}$
is unknown and so far there only exists an upper bound. Observations
of primordial gravitational waves that leave their imprint in the
$B$-polarization spectrum of the cosmic microwave radiation would allow
to determine $E_{\rm{inf}}^{\rm obs}$ in a model independent
way. 
But increasing precision will
lead to stronger bounds and eventually even to a detection that would
allow to uniquely fix $E_{\rm{inf}}^{\rm obs}$. In the following, we
will derive how $E^{\rm{obs}}_{\rm infl}$ can be expressed in terms of
observable quantities. 

The amplitude of the scalar power spectrum $A_{\rm{s}}$ is fixed by 
the measured temperature anisotropies of the CMB,
$A_{\rm{s}}\propto(\Delta T/T)^2$. For the pivot scale
$k_{*}=0.05\;\rm{Mpc}^{-1}$, the best \textsc{Planck} fit by the $\Lambda$CDM model for the scalar amplitude in the absence of tensor modes and with lensing and polarization data is \cite{P1513c4}
\bea\label{tempf}
A_{\rm{s}*} = (2.139\pm 0.063)\times 10^{-9}\label{ScalPower}
\eea
at the 68\% confidence level.

The tensor-to-scalar ratio --- to first order in the slow roll approximation --- is defined as
\bea
r:=\frac{A_{\rm{t}}}{A_{\rm{s}}}
=16\,\epsilon_{\rm{v}}=-8\,n_{\rm{t}}\,\label{TTS}.
\eea
The $B$-polarization spectrum of the CMB is produced only by tensorial
perturbations. A detection of B-modes would allow one to fix
$A_{\rm{t}\,*}$ and with (\ref{ScalPower}) also $r_{*}$. Finally, this
would allow us to determine the energy scale of inflation in a model-independent way from observations, 
\bea
(E^{\rm{obs}}_{\rm{infl}})^4:=\frac{3\,\pi^2\,M_{\rm{P}}^4}{2}\,A_{\rm{t}\,*}=\frac{3\,\pi^2\,M_{\rm{P}}^4}{2}\,A_{\rm{s}\,*}\,r_{*}.
\eea

So far, observations managed to obtain only an upper bound on $r_{*}$. 

 The current bound from \cite{PlanckBicep2} is $r_*<0.12$ at 95\% confidence level at $k_*=0.05\,\mbox{Mpc}^{-1}$. Taking the central values $A_{\rm{s}\,*}=2.2\times10^{-9}$ and saturating the bound for $r$, one obtains an upper bound for the energy scale 
\bea
 E_{\rm{infl}}^{\rm{obs}}< 1.9\times 10^{16} \,{\rm GeV}\,.
\eea
Obviously, all cosmological models have to satisfy the condition
\bea
E_{\rm{infl}}^{\rm{model}}\approx
E_{\rm{infl}}^{\rm{obs}}\label{ConsistencyCond1} 
\eea
in order not to be in conflict with observational data.

\subsection{Quantum cosmological energy scale of inflation}\label{QC}

As discussed in Sec.\ \ref{PredInQC}, we use a heuristic approach and
interpret peaks in the tunneling probability distributions as setting
the initial conditions for inflation. 
The tunneling distribution in the semiclassical limit is found to be \cite{Barvinsky:2009jd,CKS14}
\bea
 {\cal T}(\varphi):=e^{-\Gamma(\varphi)}={\rm
   exp}\left[-\frac{24\,\pi^2\,M_{\rm{P}}^4}{V(\varphi)}\right]\,.\label{ProbDistr1} 
\eea
A peak corresponds to a maximum of (\ref{ProbDistr1}).
Finding this peak is equivalent to finding the maxima of the potential $V_{\rm{\max}}:=V_{\rm{eff}}(\varphi_{\rm{\max}})$.
This leads to the simple conditions
\bea
 \frac{{\rm d}\,V(\varphi)}{{\rm
     d}\varphi}\Big|_{\varphi=\varphi_{{\rm max}}}=0,\qquad
 \frac{{\rm d}^2\,V(\varphi)}{{\rm
     d}\varphi^2}\Big|_{\varphi=\varphi_{{\rm max}}}<0\,.\label{MaxCond}  
\eea
The peak $\varphi_{\rm{max}}$ in (\ref{ProbDistr1}) corresponds to the value of $\varphi$ that selects the most probable value of $\Lambda_{\rm{eff}}=V(\varphi_{\rm{max}})/M_{\rm{P}}^2$ for which the universe starts after tunneling.
In this way, the quantum scale of inflation was obtained in \cite{Barvinsky:1995gi,Barvinsky:1996ce, Barvinsky:1998qh, Barvinsky:1998rn,Barvinsky:1999qn}.
The predictability of the tunneling distribution (\ref{ProbDistr1}) can be quantified by the sharpness of the peak at $\varphi_{\rm{max}}$,
\bea
{\cal S}:=\frac{(\Delta\varphi)^2}{E_{{\rm infl}}^{{\rm QC}}}.
\eea
Here, the variance $(\Delta\varphi)^2$ is a measure of the width of the peak, while $E^{\rm{QC}}_{\rm{infl}}$ is a measure for the height of the peak. We can get a rough estimate of the variance by fitting a normal distribution around the peak $\varphi_{\rm{max}}$. Taking $\varphi_{\rm{max}}$ as the mean value and expanding $\Gamma$ to second order around $\varphi_{\max}$, we obtain
 \bea
 (\Delta\varphi)^2:=[\Gamma''(\varphi_{\rm{max}})]^{-1}\,.
\eea
In the inflationary slow-roll regime, $\varphi\approx\rm{const}$ and the energy density is completely dominated by $V_{\rm{max}}:=V(\varphi_{\rm max})$.
Therefore, the peak value $\varphi_{\rm{max}}$ allows one to determine the energy scale of inflation as
\bea
 E_{\rm{infl}}^{\rm{QC}}:=V_{\rm{max}}^{1/4}\,\label{EInfQC}.
\eea
Demanding a consistent quantum cosmological history of the universe, beginning with the quantum creation via tunneling, we extend the consistency condition (\ref{ConsistencyCond1}) and require
\bea
 E_{\rm{infl}}^{\rm{QC}}\approx E_{\rm{infl}}^{\rm{model}}\approx E_{\rm{infl}}^{\rm{obs}}\,.\label{ConsistencyCond2}
\eea
This implies that the energy scale of the inflationary model must not
only agree with present observations but must also be of the same
order as the prediction from quantum cosmology. 

Two points deserve further discussion. First, in the presence of
multiple maxima of the effective potential, there might be several
(possibly degenerated) peaks in the probability distribution. In such
a case, the environment of these peaks, in particular the neighboring
minima, has to be investigated. 
Second, it should be noted that in the context of eternal inflation
and the landscape picture \cite{landscape1,landscape2}, such a strong
condition as $E^{\rm 
  QC}_{\rm infl}\approx E^{\rm obs}_{\rm infl}$ might not hold and one
has to resort to the somewhat weaker condition $E^{\rm QC}_{\rm
  infl}\geq E^{\rm obs}_{\rm infl}$.  Even if there was only one global
maximum of the effective potential, corresponding to the unique single
highest peak in the tunneling distribution, one must be careful in
case the effective potential also features several metastable minima
with different energy densities. Then, starting from the
global maximum, inflation could happen in a cascade-like process decaying from
higher vacua to lower ones step by step (in the following labeled by
$\varphi_{{\rm min}, n}$ where higher $n$ mean lower values of
$V(\varphi_{{\rm min},n})$) or decaying directly in some lower
metastable (or eventually even stable) vacuum. We would not be sure
whether `our inflation,' which produced our exponentially blown patch
of the universe we can observe today, was due to the initial
inflationary phase that started at the global maximum of the effective
potential or due to another inflationary period that started in a lower
metastable minimum of the effective potential. In other words, it is
logically possible that, depending on the structure of the effective
potential, the phase of inflation triggered by the quantum creation of
the universe leads to a phase of eternal inflation with
$V(\varphi_{\rm max})$. Then, in this eternally inflating universe at
some moment in time in some region of space, the inflaton field could
decay in one of the metastable vacua $\varphi_{{\rm min},n}$, starting
another phase of inflation, with a different initial condition set by
the local minimum of the effective potential $V(\varphi_{{\rm min},
  n})<V(\varphi_{\rm max})$ and this can happen several times. We
cannot really say whether the energy scale of `our' inflation is
$V(\varphi_{\rm max})$ or $V(\varphi_{{\rm min}, n})$ . Therefore, we
can realistically only demand $E^{\rm QC}_{\rm infl}\geq E^{\rm obs}_{\rm infl}$. 

In this context, it might be interesting to note that according to
\cite{Borde:2001nh} inflation can exist only eternally to the future
direction, but not to the past, assuming a universe that is on
average always expanding (not necessarily accelerated). This supports
the assumption of an initial moment of creation, in contrast to an
eternally existing inflationary universe with no beginning.

\section{Special models}\label{SpecialModels}

In the following, we will apply the general method presented in the
previous sections to two particular models of inflation: non-minimal
Higgs inflation and natural inflation; these two scenarios are among the class of
inflationary models currently favored by observational data
\cite{Ade:2013uln,Plnew}. In the following, we quote the results for the 2013 data release of \textsc{Planck}.

While natural inflation already admits a quantum cosmological analysis
at the tree level, quantum corrections are essential in non-minimal
Higgs inflation as they lead to the formation of a strict maximum in
the potential and therefore to a sharp peak in (\ref{ProbDistr1}).

\subsection{Non-minimal Higgs inflation}\label{Nmhi}

In the non-minimal Higgs inflation model \cite{Bezrukov:2007ep,Barvinsky:2008ia,Bezrukov:2008ej,DeSimone:2008ei,Barvinsky:2009fy,Bezrukov:2009db,Bezrukov:2010jz,Barvinsky:2009ii,  
  Allison:2013uaa}, the Standard Model Higgs boson and the cosmological
inflaton are identified to be one and the same scalar field $\varphi$
-- the Higgs inflaton. The other essential assumption of this model is
a strong non-minimal coupling $\xi\sim10^4-10^5$ of the Higgs inflaton
to gravity.\footnote{We note that, for an interacting scalar field
  $\varphi$, a non-mininal coupling of the form $\propto\varphi^2\,R$
  will be unavoidably induced already at the one-loop order. Even
  within an effective field theoretical approach, where higher order
  terms are supposed to be sufficiently suppressed, consistency of the
  renormalization procedure would require to include such a term
  already from the very beginning. Regarding the strength of $\xi$, we
  note that in view of the rather small mass of the discovered Higgs
  boson $M_{\rm{H}}\simeq126\;\rm{GeV}$, the condition of a large
  non-minimal coupling can be relaxed somewhat; see the discussion at
  the end of this section and, e.g., \cite{Allison:2013uaa,Bezrukov:2014ipa}.} The
interactions  
relevant for cosmology can be summarized by the 
graviton-Higgs sector of the model,
\bea
S[g,\varphi]=\int{\rm d}^4x\,g^{1/2}\left[U(\varphi)R-\frac{1}{2}\partial_{\mu}\varphi\partial^{\mu}\varphi-V(\varphi)\right].
\eea
The coupling to the Ricci scalar $R$ and the Higgs potential are given by
\bea
U(\varphi)=\frac{1}{2}\left(M_{\rm{P}}^2+\xi\varphi^2\right),\qquad V(\varphi)=\frac{\lambda}{4}\left(\varphi^2-\nu^2\right)^2.
\eea
Here, $\xi$ is the non-minimal coupling constant, $\lambda$ the quartic self-coupling, and $\nu\simeq246\; \rm{GeV}$ the symmetry breaking scale. The matter sector is given by the Standard Model interaction Lagrangian
\bea
 {\cal L}_{\rm{int}}=-\sum_{\chi}\,\frac{1}{2}\,\lambda_{\chi}\,\chi^2\,\varphi^2-\sum_{A}\,\frac{1}{2}\,g_{A}^2\,A_{\mu}^2\,\varphi^2-\sum_{\Psi}\,y_{\Psi}\varphi\,\bar{\Psi}\Psi.
\eea
The sum extends over scalar fields $\chi$, vector gauge bosons $A_{\mu}$ and fermions $\Psi$. The matter content can be restricted to the dominant contributions that come from the heavy $W^{\pm}$ and $Z$ bosons and the Yukawa top quark $q_{\rm{t}}$. Their masses follow from the relations
\bea
 m_{W^{\pm}}^2=\frac{1}{4}\,g\,\varphi^2,\quad m_{\rm{Z}}^2=\frac{1}{4}\,(g^2+g'^2)\,\varphi^2,\quad m_{\rm{t}}^2=\frac{1}{2}\,y_{\rm{t}}^2\,\varphi^2,\quad m_{\rm{H}}^2=\lambda\,(3\varphi^2-\nu^2),
\eea
with the electroweak and strong gauge couplings $g$, $g'$ and $g_{\rm{s}}$ as well as the Yukawa top quark coupling $y_{\rm{t}}$. This matter content results in essential quantum contributions to the effective potential ---the quantity that encodes the relevant information for the cosmological analysis.
Since the energy scales of the electroweak vacuum and inflation are separated by many orders of magnitude, one also needs to take into account the dependence of the coupling constants on the energy scale. In order to evaluate the coupling constants at the high energy scale of inflation, one needs to calculate the  renormalization group flow that connects the electroweak scale with the energy scale of inflation \cite{Bezrukov:2008ej,DeSimone:2008ei,Barvinsky:2009fy}. The renormalization group running of the couplings, in turn, is encoded in the beta functions which give rise to a  system of coupled ordinary differential equations that has to be solved numerically,
\bea
 \frac{d g_i(t)}{d t}
    =\beta_{g_i}, \qquad g_i=(\lambda,\xi,g,g',g_{\rm{s}},y_{\rm{t}}),\qquad\frac{dZ(t)}{d t}=\gamma Z.\label{reneq}
\eea
Here, $t=\rm{ln}\varphi/\mu$ is the logarithmic running scale and $\mu$ is an arbitrary renormalization point. 
The wave function renormalization $Z(t)$ is determined by the anomalous dimension $\gamma$ of the Higgs field.

In order to make use of the standard slow-roll formalism for the cosmological analysis, it is convenient to transform to the Einstein frame by a conformal transformation of the metric field and a redefinition of the Higgs inflaton,
\bea
  \hat g_{\mu\nu}=\frac{2U(\varphi)}
  {M_{\rm P}^2}g_{\mu\nu},\quad
  \left(\frac{d\hat\varphi}{d\varphi}\right)^2
  =\frac{M_{\rm P}^2}{2}\frac{\left(U+3U'^2\right)}{U^2},\quad\hat{V}=\frac{M_{\rm P}^2}{4}\frac{V}{U^2}\Big|_{\varphi=\hat{\varphi}}. 
\eea 
The one-loop renormalization group improved effective potential in the Einstein frame reads \cite{Barvinsky:2009fy}
\bea
  \hat{V}\simeq\frac{M_{\rm P}^4}{4}\frac{\lambda
  }{\xi^2}\,\left[1-\frac{2M_{\rm P}^2}{\xi\varphi^2}+
  \frac{\mbox{\boldmath$A_I$}}{16\pi^2}\,
  {\rm ln}\left(\frac{\varphi}{\mu}\right)\right],           
\eea
where $\mbox{\boldmath$A_I$}$ represents the inflationary anomalous
scaling \cite{Barvinsky:2009fy}
\bea
   &&{\mbox{\boldmath $A_I$}}(t) :=
    \frac3{8\lambda(t)}\Big[2g^4(t)
    +\Big(g^2(t)+g'^2(t)\Big)^2
    -16y_{\rm{t}}^4(t)\Big]-6\lambda(t).   \label{A-run}
\eea

\subsubsection{Slow-roll predictions}\label{NmhiSlowRoll}

Using the slow-roll formulas of Sec.\ \ref{ESlowRoll} for the Einstein-frame renormalization-group-improved effective potential, and taking
the derivatives with respect to the Einstein frame scalar field
$\hat{\varphi}$, one can express the the slow-roll parameters in terms
of the original Jordan frame variables \cite{Barvinsky:2009fy} 
\bea
    &&\hat\varepsilon=\frac{M_P^2}2\left(\frac1{\hat
    V}\frac{d\hat V}{d\hat\varphi}\right)^2=
    \frac43\left(
    \frac{M_P^2}{\xi\,\varphi^2}+
    \frac{\mbox{\boldmath$A_I$}}{64\pi^2}\!\right)^2,  \label{varepsilon}\\
    &&\hat\eta= \frac{M_P^2}{\hat V}
    \frac{d^2\hat V}{d\hat\varphi^2}
    =-\frac{4M_P^2}{3\xi\varphi^2}.      \label{eta}
\eea
For the expressions of the remaining cosmological parameters, it is
convenient to introduce the abbreviation 
\bea
    x:=\frac{N
    \mbox{\boldmath$A_I$}}{48\pi^2},          \label{x}
\eea
which can be interpreted as a measure of the strength of quantum corrections, resulting from $\mbox{\boldmath$A_I$}$. In this way, for the scalar amplitude one finds \cite{Barvinsky:2009fy}
\bea
    \hat{A}_{\rm s}=\frac{\lambda}{96\pi^2\xi^2\hat\varepsilon}=
    \frac{N^2}{72\pi^2}\,\frac\lambda{\xi^2}\,
    \left(\frac{e^x-1}{x\,e^x}\right)^2.       \label{HiAs}
\eea
The scalar spectral index and the tensor-to-scalar ratio are then
found to be \cite{Barvinsky:2009fy} 
\bea
  &&n_s=
  1-\frac{2}{N}\, \frac{x}{e^x-1},           \label{ns}\\
  &&r=\frac{12}{N^2}\,
  \left(\frac{x e^x}{e^x-1}\right)^2.        \label{r}
\eea
All these quantities have to be evaluated at the energy scale of of
inflation, the moment of first horizon crossing, when the inflaton
value is $\varphi_{*}$. This, in turn, means that one has to
numerically integrate the system of renormalization-group equations
(\ref{reneq}) from the electroweak vacuum $t_{\rm{ew}}\simeq0$ up to
the scale $t_{*}$, corresponding to $\varphi_{*}$, and then evaluate
all running couplings at $t_{*}$. 

Fixing the arbitrary renormalization point at the top mass scale
$\mu=M_{\rm{t}}$ and assuming a modified convention for the condition
of the end of inflation $\hat\varepsilon|_{t=t_{\rm{end}}}:=3/4$
(instead of the convention (\ref{EndOfInfl})), the times $t_*$
and $t_{\rm{end}}$ can be determined via the relation
$\varphi_{*/\rm{end}}=M_t\exp(t_{*/\rm{end}})$ and the estimate for
the number of e-folds \cite{Barvinsky:2008ia}, 
\bea
    N_{*}\simeq\frac34\,\frac{\xi(t_{*})}{M_{\rm P}^2}\,
    (\varphi_{*}^2-\varphi_{\rm end}^2)~.                               \label{NBezShap}
\eea
In the large $\xi$ approximation, these times read
\cite{Barvinsky:2009fy} 
\bea
    &&t_{*} = \ln \frac{M_{\rm P}}{M_{\rm t}}+
    \frac12 \ln\frac{4N}{3\xi_{*}}
    +\frac12
    \ln\frac{\exp x_{*}-1}{x_{*}},\\\label{relation1}
    &&t_{\rm end} = \ln \frac{M_{\rm P}}{M_{\rm t}}
    +\frac12 \ln\frac4{3\xi_{\rm end}}.                  \label{end1}
\eea
While we have taken into account a running $\xi$, numerically the
running is very slow, i.e. $\xi(t_{\rm
  ew})\simeq\xi(t_{*})$. Nevertheless, in view of the fact that there
is no initial condition for $\xi(t_{\rm ew})$, one has to impose a
`final condition' $\xi(t_{*})$, determined by the correct
normalization of the scalar amplitude (\ref{HiAs}) evaluated at
$t_{*}$, i.e. $A_{\rm s *}\propto\lambda_{*}/\xi_{*}^2\propto(\Delta
T/T)^2\sim10^{-10}$. 
Note that for $N=50\div 60$, the duration of inflation in terms of the
logarithmic scale $t$ is numerically very short
$t_{*}-t_{\rm{end}}\simeq2$ compared to the post-inflationary running
$t_{\rm end}-t_{\rm ew}\simeq35$ \cite{Barvinsky:2009fy}.

The numerical predictions for (\ref{ns}) and (\ref{r}) depend on the
details of the renormalization-group flow and are, in particular, very
sensitive to the initial conditions at the electroweak scale.  A more precise analysis including beta
functions up to two and three loops has become available, see
e.g. \cite{Bezrukov:2009db,Bezrukov:2010jz} and, since the discovery
of the Higgs boson, also the initial conditions at the electroweak
scale (in particular the top mass $M_{\rm t}$) are known to a higher
precision. Another aspect is
connected to the rather light value of the measured Higgs mass
$M_{\rm{H}}\simeq126\,\rm{GeV}$. As a consequence, $\lambda(t_{\rm
  in})$ can be very small by itself at the energy scale of inflation
and therefore allows for much smaller non-minimal couplings
$\xi(t_{\rm in})$; see e.g. a discussion in
\cite{Allison:2013uaa,Bezrukov:2014ipa}. For certain initial values
the renormalization group flow can even drive $\lambda$ to negative values and therefore
lead to an unstable (or meta-stable) vacuum, see e.g. \cite{Buttazzo:2013uya}.
In general, the precise  inflationary predictions of this model are very sensible to small
changes in initial values for the Standard Model masses.

\subsubsection{Quantum cosmological predictions}\label{NmhiQC}

Following the general method presented in Sec. \ref{QC}, the tunneling
amplitude for the non-minimal Higgs inflation model is given by
\cite{Barvinsky:2009jd} 
\bea
\Gamma(\varphi)=24\pi^2\frac{M_{\rm P}^4}{\hat V(\varphi)}
\simeq 96\pi^2\frac{\xi^2}
\lambda\left(1+\frac{2M_{\rm P}^2}{\xi Z^2\varphi^2}\right).
\eea
The peak position $\varphi_{\rm{max}}$ is determined by the extrema
\bea
  \varphi\frac{d\Gamma}{d\varphi}=\frac{d\Gamma}{dt}
  =-\frac{6\xi^2}
  \lambda\left(\mbox{\boldmath$A_I$}
  +\frac{64\pi^2M_{\rm P}^2}{\xi Z^2\varphi^2}\right)=0,
\eea
The solution of this condition in terms of the probability peak reads
\cite{Barvinsky:2009jd} 
\bea
    \varphi^2_{\rm{max}}=
    \left.-\frac{64\pi^2M_{\rm P}^2}{\xi \mbox{\boldmath$A_I$}Z^2}
    \,\right|_{\;t=t_{\rm{max}}}.                               \label{root}
\eea
 The peak is very narrow, as can be estimated by the sharpness 
\bea
    {\cal S}=\frac{(\Delta\varphi)^2}{E^{\rm QC}_{\rm inf}}
    \simeq\frac{\frac{d^2\Gamma(t)}{dt^2}}{\hat{V}(t)}\Bigg|_{t=t_{\rm max}}=-\left.\frac\lambda{12\xi^2}
    \frac1{\mbox{\boldmath$A_I$}}\right|_{t=t_{\rm \max}}\sim
    10^{-10}.
\eea
In view of $\mbox{\boldmath$A_I$}(t_{\rm
max})\simeq\mbox{\boldmath$A_I$}(t_{\rm end})$, the point of the horizon
crossing $\varphi_{*}$ for the pivot scale $k$, chosen to correspond to $N=60$, is very close to the point of quantum creation $\varphi_{\rm{max}}$. Their ratio for different modes, corresponding to different $N$, therefore takes the form \cite{Barvinsky:2009jd}
\bea
  &&\frac{\varphi^2_{*}}{\varphi_{\rm max}^2}=
  1-\exp\left[-N\frac{|\mbox{\boldmath$A_I$}
  (t_{\rm end})|}{48\pi^2}\right].            \label{phi0phiin}
\eea
Thus, (\ref{phi0phiin}) indicates that, for wavelengths longer than the
pivotal one, the 
instant of horizon crossing approaches the moment of `creation' of
the Universe, but it is always posterior to it ($\varphi_{\rm max}>\varphi_*$), as required for a consistent quantum cosmological history of the universe. 

\subsection{Natural inflation}\label{Nati}

Another inflationary model which is in agreement with the Planck data
is that of natural inflation \cite{Freese:1990rb}. 
The inflaton potential for natural inflation reads 
\bea
 V=\Lambda^4\,\left[1+\cos\left(\varphi/f\right)\right]\,.\label{NatInfPot}
\eea
In this model, $\varphi$ is supposed to be a pseudo Nambu--Goldstone boson taking values
on a circle with radius $f$ and angle $\varphi/f\in[0,\,2\,\pi)$ \cite{Freese:1990rb}. The two constants $\Lambda$ and $f$
determine the height and the slope of the potential and have physical
dimension of mass in natural units. The interpretation of $\varphi$ as
a pseudo Nambu--Goldstone field suggests that $f=O(M_{\rm{P}})$ and
$\Lambda\approx M_{\rm{GUT}}\sim 10^{16}$ GeV.

\subsubsection{Slow-roll predictions}\label{NatiSlowRoll}

The cosmological parameters in the inflationary slow-roll analysis can
again be derived from the general expressions in
Sec. \ref{ESlowRoll}. From (\ref{SlowRollParameters}), the first two
slow-roll parameters read 
\bea\label{srpa}
 \epsilon_{\rm{v}}=\frac{M_{\rm{P}}^2}{2\,
   f^2}\,\tan^2\left[\varphi/(2\, f)\right],\qquad 
 \eta_{\rm{v}}=-\frac{M_{\rm{P}}^2\, \cos(\varphi/f)}{f^2\,
   \left[1 + \cos(\varphi/f)\right]}. 
\eea
The scalar spectral index and the tensor-to-scalar ratio then take the form
\bea
n_{\rm{s}}=-\frac{M_{\rm{P}}^2}{f^2}\,\frac{3-\cos(\varphi/f)}{1+\cos(\varphi/f)},\qquad\label{SpectralIndex}
r=\frac{8\,M_{\rm{P}}^2}{ f^2}\,\tan^2\left[\varphi/(2\, f)\right]\,.\label{TensorToScalarRatio}
\eea
All cosmological observables must again be evaluated at $\varphi_{*}$,
the field value that corresponds to the moment where the pivot mode
$k_{*}$ first crosses the Hubble scale. 
The number of e-folds $N_*$ connecting the end of inflation
$\varphi_{\rm{end}}$ with the value $\varphi_{*}$ is 
\bea
 N_{*}=\frac{2\,f^2}{M_{\rm{P}}^2}\,\ln\left[\frac{\sin\left(\frac{\varphi_{\rm{end}}}{2\,f}\right)}{\sin\left(\frac{\varphi_{*}}{2\,f}\right)}\right]\,.\label{EFoldsNatural}
\eea
The value $\varphi_{\rm{end}}$ that determines the upper integration bound in (\ref{EFoldsNatural}) is determined to be
\bea
 \varphi_{\rm{end}}=2\, f\, {\rm arctan} (\sqrt{2}\, f/M_{\rm{P}})\,.\label{PhiEnd}
\eea
Inserting (\ref{PhiEnd}) in (\ref{EFoldsNatural}), solving for $\varphi_{*}$ and parametrizing $f$ 
in units of $M_{\rm{P}}$, we find
\bea
 \varphi_{*}={}2 M_{\rm{P}}\,\alpha\, {\rm arcsin}\left(\frac{\alpha\,e^{-N_{*}/2 \alpha^2}}{\sqrt{
  1/2 + \alpha^2}}\right),\label{phiStar}
\eea
where $\alpha:=f/M_{\rm{P}}$.
Evaluating the potential (\ref{NatInfPot}) at $\varphi_{*}$ yields
\bea
 V(\varphi_{*})=2\,\Lambda^4\,\left[1-\delta_{V}(\alpha,N_{*})\right]\,,
\eea
where, following \cite{CKS14}, we have defined
\bea
 \delta_{V}(N_{*},\alpha):=\frac{2\,e^{-N_{*}/\alpha^2}\,\alpha^2}{1+2\,\alpha^2}\,.
\eea
\begin{figure}[ht!]
\begin{center}
\includegraphics[scale=0.8]{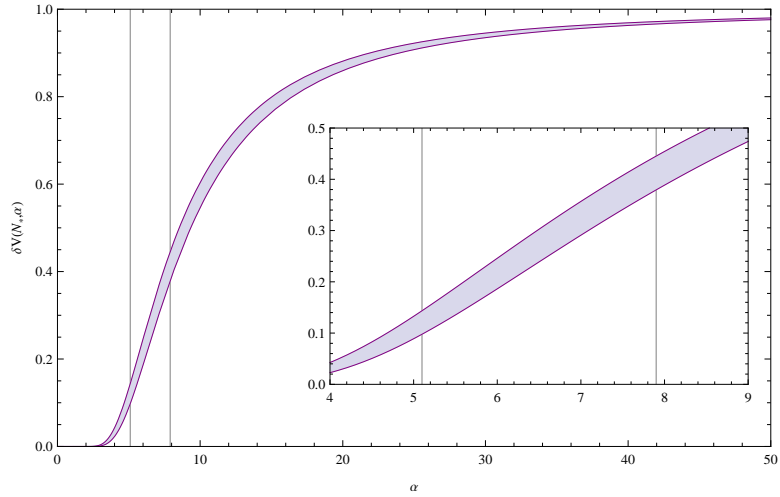}
\caption{\label{fig1} The function $\delta_V(N_{*},\alpha)$ as a
  function of $\alpha$ for values of $N_{*}\in[50,60]$, taken from
  \cite{CKS14}. The upper line corresponds to $N_{*}=50$, the lower
  line to $N_{*}=60$. The inset shows the region with $\alpha$ in the
  $68\%$ CL range $5.1< \alpha < 7.9$ (see (\ref{PlanckDataRange})). 
}
\end{center}
\end{figure}

The expressions for $n_{\rm{s}}$ and $r$ evaluated at $\varphi_{*}$ can be expressed in terms of $\delta_V$ and $\alpha$:
\bea
n_{\rm{s}\,*}={}1+\frac{1}{\alpha^2}\,\frac{\delta_V(N_{*},\,\alpha)+1}{\delta_V(N_{*},\alpha)-1}\,,\qquad r_{*}={}\frac{8}{\alpha^2}\,\frac{\,\delta_V(N_{*},\,\alpha)}{1-\,\delta_V(N_{*},\,\alpha)}\,.
 \eea
For $N_{*}=60$, \textsc{Planck} 2013 data \cite{Ade:2013uln} constrain $\alpha$ to lie in the interval \cite{Tsujikawa:2013ila}
\bea
 5.1<\alpha<7.9\quad(68\%\,\rm{CL})\,.\label{PlanckDataRange}
\eea

\subsubsection{Quantum cosmological predictions}\label{NatiQC}

Following again the general algorithm of Sec. \ref{QC}, we first
have to calculate the extrema of (\ref{NatInfPot}), 
\bea
\frac{\rm{d} V}{\rm{d}\varphi}\Big|_{\varphi=\varphi_{\rm{ext}}}=-\frac{\Lambda^4}{f}\,\sin\left(\varphi_{\rm{ext}}/f\right)=0\,.
\eea
If $\varphi_{\rm{ext}}$ is a maximum, peak values correspond to
\bea
\varphi_{\rm{max}}:=2\,\pi nf\,.
\eea
The potential at $\varphi_{\rm{max}}$ has the value
\bea
V_{\rm{max}}=2\,\Lambda^4\,.\label{VatMax}
\eea
With the width $\Delta\varphi$ of the distribution given by
\bea
 (\Delta\varphi)^2= \left.\frac{1}{\Gamma''}\right|_{\varphi=\varphi_{\rm{max}}}=\frac{1}{6\,\pi^2}\,\frac{f^2\,\Lambda^4}{M_{\rm{P}}^4}\,,\label{varianceGaussFit}
\eea
the sharpness of the peak $\varphi_{\rm{max}}$ is estimated as
\bea
 {\cal S}=\frac{(\Delta\varphi)^2}{V_{\rm{max}}^{1/2}}
 \approx\frac{1}{6\,\pi^2}\,\frac{f^2\,\Lambda^2}{M_{\rm{P}}^4}\sim\frac{\Lambda^2}{M_{\rm{P}}^2}\sim 10^{-4}\,,
\eea
where we have used $f\sim M_{\rm{P}}$ and $E^{\rm{QC}}_{\rm{inf}}\sim\Lambda$, according to (\ref{EInfQC}) and (\ref{VatMax}).

\begin{figure}[ht!]
\begin{center}
\includegraphics[scale=0.9]{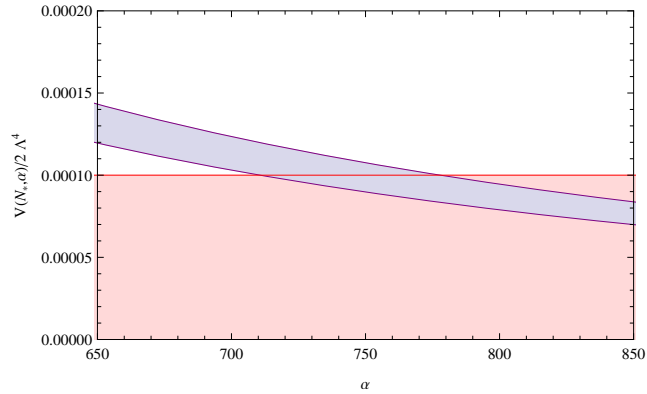}
\caption{\label{fig2} A zoomed-in region of the function $V(N_{*},\alpha)/(2\,\Lambda^4)=1-\delta_V$ as a function of $\alpha$ for values of $N_{*}\in[50,60]$, taken from \cite{CKS14}. The upper purple line corresponds to $N_{*}=60$, the lower purple line to $N_{*}=50$. The lower area, colored in light red (in black-and-white printing: light gray), corresponds to the region where $E_{\rm{inf}}^{\rm{model}}<10^{-1}\,E_{\rm{inf}}^{\rm QC}$.
}

\end{center}
\end{figure}

As can be inferred from Fig.~\ref{fig2}, it was shown in \cite{CKS14}
that the requirement of a deviation from the approximate consistency
requirement (\ref{ConsistencyCond2}) of not more than one order of
magnitude leads to the following constraint for the
parameter $\alpha$:  
\bea
\alpha \lesssim 710\;\; {\rm for}\;\;N_{*}=50 \qquad {\rm and}\qquad \alpha \lesssim 780\;\; {\rm for}\;\; N_{*}=60.\label{QCPredAlpha}
\eea

Although a quantum cosmological bound on $\alpha$ derived in this way
depends on the allowed tolerance for a violation of
(\ref{ConsistencyCond2}), this bound is clearly not as restrictive as
the constraints on $\alpha$ coming from the comparison with the
observational constraints of the spectral index and the
tensor-to-scalar ratio by \textsc{Planck}. 
As shown in Fig.~\ref{fig1}, in this range $\delta_V\approx 0.1\div
0.5$ is still small enough to respect the approximate condition
(\ref{ConsistencyCond2}) to the tolerated accuracy. 

Thus, consistency of classical inflationary predictions
with observational data (\ref{PlanckDataRange}) result in a much
sharper bound $\alpha\sim O(10)$ far below the threshold
$\alpha\approx 700$. We can therefore conclude that to a good
approximation no conflict with the a quantum origin or our universe
does arise in natural inflation, since the consistency condition is
satisfied for all experimentally allowed values of $\alpha$.

\section{Conclusions and outlook}\label{Con}

In this contribution, we have presented a general method that allows one
to derive predictions from quantum cosmology by assuming a consistent
history of our universe from its initial quantum creation up to its
present state. We have restricted our analysis here to the tunneling
scenario, but the method can also be extended to other
quantum initial conditions such as the no-boundary condition, 
although this condition does not lead naturally to inflationary
initial conditions. 

We have in detail investigated two particular models of inflation:
non-minimal Higgs inflation
\cite{Bezrukov:2007ep,Barvinsky:2008ia,Bezrukov:2008ej,DeSimone:2008ei,Barvinsky:2009fy,Bezrukov:2009db,Bezrukov:2010jz,Barvinsky:2009ii,  
  Allison:2013uaa} and natural inflation \cite{Freese:1990rb}. We have 
found that both models allow for a consistent cosmic history starting
from a quantum tunneling process. 
  
In principle, all inflationary single-field models favored by recent \textsc{Planck} data can be summarized by the general class of scalar-tensor theories with the action
\bea
 S=\int{\rm d}^4x\,\sqrt{g}\,\left[U(\varphi)\,R-\frac{G(\varphi)}{2}\,(\nabla\varphi)^2-V(\varphi)\right]\label{ActionScalarTensor1}\,,
\eea
and the method presented here could, in principle, be applied also to this general class parametrized in terms of the arbitrary functions $U(\varphi)$, $G(\varphi)$ and $V(\varphi)$. 
As has been discussed in the context of non-minimal Higgs inflation, quantum corrections can become important and modify the shape and the location of the extrema of the effective potential.
The one-loop divergences for the general action (\ref{ActionScalarTensor1}), necessary for renormalization, were obtained in \cite{Steinwachs:2011zs} in a closed form for an even more general setup of a symmetric $O(N)$ invariant multiple of scalar fields.

Finally, a note regarding the parametrization dependence of these
quantum corrections is in order. 
While in the transition from the Jordan to the Einstein frame
parametrizations leads to equivalent formulations at tree level, in
the usual quantum field theoretical formalism such a field
transformation will in general induce an off-shell parametrization
dependence of the effective action
\cite{Vilkovisky:1984st,Steinwachs:2013tr,Kamenshchik:2014waa}. In
\cite{Vilkovisky:1984st}, a geometric approach to the effective action
was suggested to overcome the problem with non-covariance (with
respect to the configuration space of field) of the ordinary
formalism. Recently, this idea has been adopted in \cite{Moss:2014nya}
in the context of non-minimal Higgs inflation. 
\vskip 5mm




\begin{thebibliography}{99}

\bibitem{Mukhanov:book}
Mukhanov V 2012 {\em Physical Foundations of Cosmology} third edition (Cambridge: Cambridge University Press)

\bibitem{Martin15}
Martin J 2015 The observational status of cosmic inflation after Planck {\em Preprint} astro-ph.CO/1502.05733

\bibitem{OUP}
Kiefer C 2012 {\em Quantum Gravity} third edition (Oxford: Oxford University Press)
  
\bibitem{Barvinsky:2009jd}
Barvinsky A O, Kamenshchik A Yu, Kiefer C and Steinwachs C F 2010 Tunneling cosmological state revisited: Origin of inflation with a non-minimally coupled Standard Model Higgs inflaton {\em Phys. Rev.} D {\bf 81} 043530 ({\em Preprint} hep-th/0911.1408) 
  
\bibitem{CKS14}
Calcagni G, Kiefer C and Steinwachs C F 2014 Quantum cosmological consistency condition for inflation {\em J. Cosmol. Astropart. Phys.} {\bf10} 026

\bibitem{Kiefer:2013jqa}
Kiefer C 2013 Conceptual problems in quantum gravity and quantum cosmology {\em ISRN Math. Phys.} {\bf 2013} 509316 ({\em Preprint} gr-qc/1401.3578)

\bibitem{Hartle87}
Hartle J B 1987 {\em Prediction in quantum cosmology} (Gravitation in astrophysics) ed B Carter and J B Hartle (New York: Plenum Press) pp 329--60

\bibitem{Halliwell91}
Halliwell J J 1991 {\em Introductory lectures on quantum cosmology} (Quantum cosmology and baby universes) ed S Coleman, J B Hartle, T Piran and S Weinberg (Singapore: World Scientific) pp 159--243

\bibitem{deco}
Joos E, Zeh H D, Kiefer C, Giulini D, Kupsch J and Stamatescu I O 2003 {\em Decoherence and the Appearance of a Classical World in Quantum Theory} second edition (Berlin: Springer)

\bibitem{Schlosshauer}
Schlosshauer M 2007 {\em Decoherence and the quantum-to-classical transition} (Berlin: Springer)

\bibitem{Zeh07}
Zeh H D 2007 {\em The Physical Basis of the Direction of Time} fifth edition (Berlin: Springer)

\bibitem{Vilenkin89}
Vilenkin A 1989 Interpretation of the wave function of the universe {\em Phys. Rev.} D {\bf 39} 1116  

\bibitem{Zurek05}
Zurek W H 2005 Probabilities from entanglement, Born's rule $p_k=\vert\psi_k\vert^2$ from envariance {\em Physical Review} A {\bf 71} 052105

\bibitem{HaPa86}
Hawking S W and Page D N 1986 Operator ordering and the flatness of the Universe {\em Nucl. Phys.} B {\bf 264} 185

\bibitem{BK90}
Barvinsky A O and Kamenshchik A Yu 1990 One loop quantum cosmology: the normalizability of the Hartle--Hawking wave function and the probability of inflation {\em Class. Quantum Grav.} {\bf 7} L181

\bibitem{Conradi}
Conradi H D 1998  Tunneling of macroscopic universes {\em Int. J. Mod. Phys.} D {\bf 7} 189

\bibitem{Vilenkin15}
Mithani A T and Vilenkin A 2015 Tunneling decay rate in quantum cosmology {\em Preprint} hep-th/1503.00400

\bibitem{Page}
Page D N 2014 Spacetime average density (SAD) cosmological measures {\em J. Cosmol. Astropart. Phys.} {\bf11} 038 
  
\bibitem{PlanckBicep2}
Ade P A R {\em et al.} 2015 A joint analysis of BICEP2/Keck Array and Planck data {\em Preprint} astro-ph.CO/1502.00612 

\bibitem{P1513c4}
Ade P A R {\em et al.} 2015 Planck 2015 results. XIII. Cosmological parameters {\em Preprint} astro-ph.CO/1502.01589

\bibitem{Barvinsky:1995gi}
Barvinsky A O and Kamenshchik A Yu 1996 Quantum origin of the early universe and the energy scale of inflation {\em Int. J. Mod. Phys.} D {\bf 5} 825-844 ({\em Preprint} gr-qc/9510032) 

\bibitem{Barvinsky:1996ce}
Barvinsky A O, Kamenshchik A Yu and Mishakov I V 1997 Quantum origin of the early inflationary universe {\em Nucl. Phys.} B {\bf 491} 387-426 ({\em Preprint} gr-qc/9612004) 

\bibitem{Barvinsky:1998qh}
Barvinsky A O 1999 Open inflation from quantum cosmology with a strong nonminimal coupling {\em Nucl. Phys.} B {\bf 561} 159-187 ({\em Preprint} gr-qc/9812058) 

\bibitem{Barvinsky:1998rn}
Barvinsky A O and Kamenshchik A Yu 1998 Effective equations of motion and initial conditions for inflation in quantum cosmology {\em Nucl. Phys.} B {\bf 532} 339-360 ({\em Preprint} hep-th/9803052)

\bibitem{Barvinsky:1999qn}
Barvinsky A O, Kamenshchik A Yu and Kiefer C 1990 Origin of the inflationary universe {\em Mod. Phys. Lett.} A {\bf 14} 1083 ({\em Preprint} gr-qc/9905098)

\bibitem{landscape1}
Carr B (ed) 2007 {\em Universe or Multiverse?} (Cambridge: Cambridge University Press)

\bibitem{landscape2}
Douglas M R 2003 The statistics of string/M theory vacua {\em J. High Energy Phys.} {\bf05} 046 

\bibitem{Borde:2001nh}
Borde A, Guth A H and Vilenkin A 2003 Inflationary space-times are incomplete in past directions {\em Phys. Rev. Lett.} {\bf 90} 151301 ({\em Preprint} gr-qc/0110012)  
  
\bibitem{Ade:2013uln}
Ade P A R {\em et al.} Planck Collaboration 2013 Planck 2013 results. XXII. Constraints on inflation {\em Preprint} astro-ph.CO/1303.5082

\bibitem{Plnew} Ade P A R {\em et al.} Planck Collaboration 2015 Planck 2015. XX. Constraints on inflation {\em Preprint} astro-ph.CO/1502.02114
  
\bibitem{Bezrukov:2007ep}
Bezrukov F L and Shaposhnikov M 2008 The Standard Model Higgs boson as the inflaton {\em Phys. Lett. B} {\bf 659} 703-706 ({\em Preprint} hep-th/0710.3755) 

\bibitem{Barvinsky:2008ia}
Barvinsky A O, Kamenshchik A Yu and Starobinsky A A 2008 Inflation scenario via the Standard Model Higgs boson and LHC {\em J. Cosmol. Astropart. Phys} {\bf 11} 021 ({\em Preprint} hep-ph/0809.2104) 

\bibitem{Bezrukov:2008ej}
Bezrukov F L, Magnin A and Shaposhnikov M 2009 Standard Model Higgs boson mass from inflation {\em Phys. Lett. B} {\bf 675} 88-92 ({\em Preprint} hep-ph/0812.4950) 

\bibitem{DeSimone:2008ei}
De Simone A, Hertzberg M P and Wilczek F 2009 Running inflation in the Standard Model {\em Phys. Lett. B} {\bf 678} 1-8 ({\em Preprint} hep-ph/0812.4946) 
  
\bibitem{Barvinsky:2009fy}
Barvinsky A O, Kamenshchik A Yu, Kiefer C, Starobinsky A A and Steinwachs C 2009 Asymptotic freedom in inflationary cosmology with a non-minimally coupled Higgs field {\em J. Cosmol. Astropart. Phys} {\bf 12} 003 ({\em Preprint} hep-ph/0904.1698) 

\bibitem{Bezrukov:2009db}
Bezrukov F and Shaposhnikov M 2009 Standard Model Higgs boson mass from inflation: two loop analysis {\em J. High Energy Phys.} {\bf 07} 089 ({\em Preprint} hep-ph/0904.1537) 

\bibitem{Bezrukov:2010jz}
Bezrukov F, Magnin A, Shaposhnikov M and Sibiryakov S 2011 Higgs inflation: consistency and generalisations {\em J. High Energy Phys.} {\bf 01} 016 ({\em Preprint} hep-ph/1008.5157) 

\bibitem{Barvinsky:2009ii}
Barvinsky A O, Kamenshchik A Yu, Kiefer C, Starobinsky A A and Steinwachs C F 2012 Higgs boson, renormalization group, and naturalness in cosmology {\em Eur. Phys. J. C} {\bf72} 2219 ({\em Preprint} hep-ph/0910.1041)   
  
\bibitem{Allison:2013uaa}
Allison K 2014 Higgs xi-inflation for the 125-126 GeV Higgs: a two-loop analysis {\em J. High Energy Phys.} {\bf 1402} 040 ({\em Preprint} hep-ph/1306.6931) 

\bibitem{Bezrukov:2014ipa}
Bezrukov F, Rubio J and Shaposhnikov M 2014 Living beyond the edge: Higgs inflation and vacuum metastability {\em Preprint} hep-ph/1412.3811  
  
\bibitem{Buttazzo:2013uya}
Buttazzo D, Degrassi G, Giardino P P, Giudice G F, Sala F, Salvio A and Strumia A 2013 Investigating the near-criticality of the Higgs boson {\em J. High Energy Phys.} {\bf 1312} 089 ({\em Preprint} hep-ph/1307.3536)  
  
\bibitem{Freese:1990rb}
Freese L, Frieman J A, Olinto A V 1990 Natural inflation with pseudo-Nambu--Goldstone bosons {\em Phys. Rev. Lett.} {\bf 65} 3233-3236   

\bibitem{Tsujikawa:2013ila}
Tsujikawa S, Ohashi J, Kuroyanagi S and De Felice A 2013 Planck constraints on single-field inflation {\em Phys. Rev.} D {\bf 88} 023529 ({\em Preprint} astro-ph.CO/1305.3044)
  
\bibitem{Steinwachs:2011zs}
Steinwachs C F and Kamenshchik A Yu 2011 One-loop divergences for gravity non-minimally coupled to a multiplet of scalar fields: calculation in the Jordan frame. I. The main results {\em Phys. Rev.} D {\bf 84} 024026 ({\em Preprint} gr-qc/1101.5047) 

\bibitem{Vilkovisky:1984st}
Vilkovisky G A 1984 The unique effective action in quantum field theory {\em Nucl.Phys.} B {\bf 234 } 125-137  
  
\bibitem{Steinwachs:2013tr}
Steinwachs C F and Kamenshchik A Yu 2012 Non-minimal Higgs inflation and frame dependence in cosmology {\em AIP Conf.Proc.}  {\bf 1514} 161-164 ({\em Preprint} gr-qc/1301.5543)  
  
\bibitem{Kamenshchik:2014waa}
Kamenshchik A Yu and Steinwachs C F 2014 Frame dependence of quantum corrections in cosmology ({\em Preprint} gr-qc/1408.5769)

\bibitem{Moss:2014nya}
Moss I G 2014 Covariant one-loop quantum gravity and Higgs inflation {\em Preprint} hep-th/1409.2108
\end{thebibliography}
\end{document}